\documentclass[12pt]{article}
\usepackage{amsfonts}
\usepackage[dvips]{graphics}
\newcommand{\rem}[1]{{\bf Remark:}}

\def\QED{{\hspace*{\fill}{\vrule height .5ex width 1ex }\quad}
    \vskip 0pt plus20pt}
\newcommand{\be}{\begin{equation}}
\newcommand{\ee}{\end{equation}}
\newcommand{\bea}{\begin{eqnarray}}
\newcommand{\eea}{\end{eqnarray}}
\newcommand{\beann}{\begin{eqnarray*}}
\newcommand{\eeann}{\end{eqnarray*}}
\begin{document}


\vspace{20pt}

\begin{center}
{\LARGE \bf Kinematics on oblique axes\\[27pt]}
{\large \bf Oscar Bolina \\[10pt]}
{\large  Departamento de F\'{\i}sica-Matem\'atica\\[3pt]
Universidade de S\~ao Paulo\\[3pt]
Caixa Postal 66318 \\
S\~ao Paulo 05315-970 Brasil\\[5pt]}
{\bf E-mail}; \normalsize{bolina@if.usp.br}\\[30pt]
\end{center}
\vskip .3 cm
\noindent
{\abstract
\noindent
{\normalsize We solve a difficult problem involving
velocity and acceleration components along oblique 
axes, and propose two problems of central 
force motion to be solved using oblique axes.
 } 
\[
\]
}
\vfill
\hrule width2truein \smallskip {\baselineskip=10pt \noindent Copyright
\copyright\ 2001 by the author. Reproduction of this article in its
entirety is permitted for non-commercial purposes.\par }

\section{Introduction}
Any vector quantity can be resolved into components according to 
the parallelogram law, both by rectangular and oblique resolution.
\newline
In rectangular coordinates, one component is perpendicular to the other.
In oblique coordinates, one component has a projection on the other. 
We have to take this projection into account when finding
velocity and acceleration components along these axes.
\newline
Because of this, problems involving oblique axes are very 
difficult. However, once one realizes that a problem
requires oblique axes, the solution is not in general 
that hard, although the derivation of the kinematics on oblique 
axes is somewhat disgusting. 
\vskip .3 cm
\noindent
\section{The velocity components}
\begin{figure}[t]
\begin{center}
\resizebox{!}{7.4 truecm}{\includegraphics{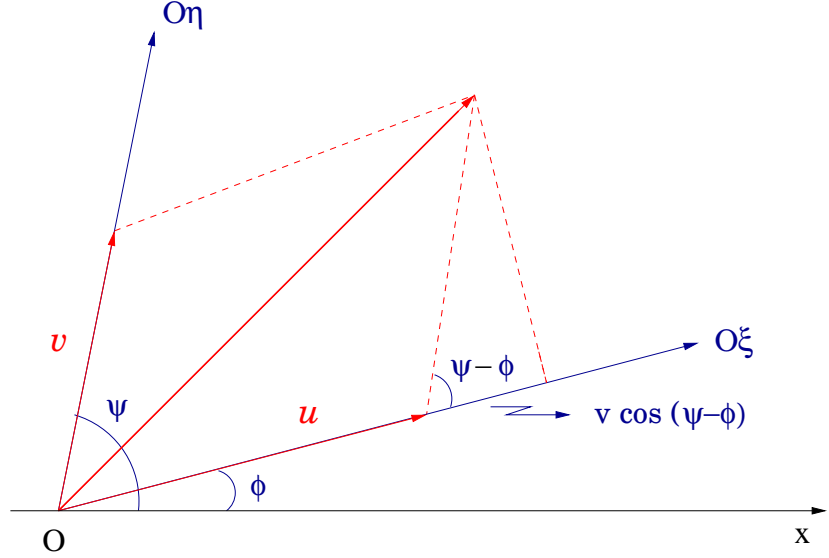}}
\vskip .2 cm
\parbox{13truecm}{\caption{\baselineskip=16 pt\small\label{oblique}
Oblique coordinate system and the components of the velocity 
along oblique axes. Each component has a projection on the other axis.
}
}
\end{center}
\end{figure}

\begin{figure}[t]
\begin{center}
\resizebox{!}{8 truecm}{\includegraphics{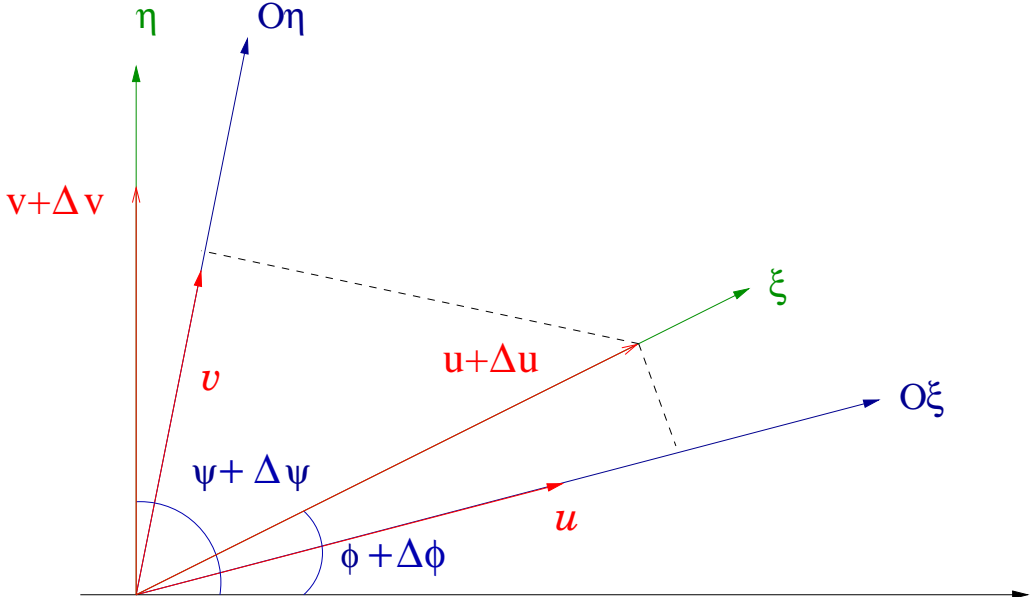}}
\vskip .2 cm
\parbox{13truecm}{\caption{\baselineskip=16 pt\small\label{oblique3}
The component of the velocity $u+\Delta u$ along the new axis $\xi$ 
is projected on the old axis $O \xi$. 
}
}
\end{center}
\end{figure}

\vskip .2 cm
\noindent
Consider the motion of a particle in a plane. Suppose that the 
geometry of the motion is such that the velocity of the particle 
is more conveniently referred to two oblique axes $O\xi$ and $O\eta$
which make angles $\phi$ and $\psi$ respectively with a fixed direction
$Ox$ in the plane, as shown in Fig. \ref{oblique}. Theses angles may vary
arbitrarily with time as the particle moves.
\vskip .1 cm
\noindent
Suppose that at the time $t$ the components of the velocity in the
directions $O\xi$ and $O\eta$ are $u$ and $v$, respectively.
The {\it perpendicular} projections of these components along 
$O\xi$ and $O\eta$ are, respectively,
\vskip .4 cm
\noindent
\be\label{1}
u+v\cos(\psi-\phi) \;\;\;\;\;\;\;\; {\rm and}
\;\;\;\;\;\;\;\; v+u\cos(\psi-\phi),
\ee
\vskip .2 cm
\noindent
as shown in the figure for the projection on the $O\xi$ axis only.
\vskip .2 cm
\noindent
At the time $t+\Delta t$ the axes $O\xi$ and $O\eta$ take 
the positions  $\xi$ and $\eta$, as shown in Fig. \ref{oblique3},
with $\xi O\xi=\Delta\phi$ and $\eta O\eta=\Delta\psi$. 
\noindent
\vskip .1 cm
\noindent
Let the components of the velocity along these axes at this time be
$u+\Delta u$ and $v+\Delta v$. The perpendicular 
projections of these components along the axes $O\xi$
and $O\eta$ are, respectively,
\[
(u+\Delta u)\cos\Delta\phi+(v+\Delta v)\cos(\psi-\phi+\Delta\psi)
\]
and
\be\label{2} 
(v+\Delta v)\cos\Delta\psi+(u+\Delta u)\cos(\psi-\phi-\Delta\phi).
\ee
\vskip .2 cm
\noindent
By taking the difference between the projections (\ref{2}) and
(\ref{1}) of the velocities along the axes $O\xi$ and $O\eta$ 
at the corresponding times $t+\Delta t$ and $t$, dividing the 
result by $\Delta t$, and letting $\Delta t$ go to zero, we 
obtain the projections of the acceleration along
those axes at the time $t$: 
\be
\dot{u}+\dot{v}\cos(\psi-\phi)-v\dot{\psi}\sin(\psi-\phi)
\ee
and
\be
\dot{v}+\dot{u}\cos(\psi-\phi)+u\dot{\phi}\sin(\psi-\phi),
\ee
\vskip .2 cm
\noindent
where $\dot{u}$ is the limiting value of ${\Delta
u}/{\Delta t}$, when $\Delta t$ approaches zero.
\vskip .2 cm
\noindent
\section{The acceleration components}
\vskip .2 cm
\noindent
\noindent
Now let $a_{\xi}$ and $a_{\eta}$ represent the components of the
acceleration of the particle along $O\xi$ and $O\eta$ at
the time $t$. The {\it same} relationship (\ref{1}) for 
velocities hold for accelerations. 
Thus, the perpendicular projections of the components along the axes
$O\xi$ and $O\eta$ are
\be\label{4}
a_{\xi}+a_{\eta}\cos(\psi-\phi) \;\;\;\;\;\;\;\; {\rm and} 
\;\;\;\;\;\;\;\; a_{\eta}+a_{\xi}\cos(\psi-\phi).
\ee
On equating (\ref{2}) and (\ref{4}) we obtain
\[
a_{\xi}+a_{\eta}\cos(\psi-\phi)=
\dot{u}+\dot{v}\cos(\psi-\phi)-v\dot{\psi}\sin(\psi-\phi)
\]
and 
\be\label{5}
a_{\eta}+a_{\xi}\cos(\psi-\phi)=
\dot{v}+\dot{u}\cos(\psi-\phi)+u\dot{\phi}\sin(\psi-\phi),
\ee
from which we can solve for $a_{\xi}$ and $a_{\eta}$. 
\vskip .2 cm
\noindent
The first two terms in equations (\ref{5}) are the rates of change of
the projections (\ref{1}) along fixed axes. The last terms are the
consequence of the motion of the axes themselves. (Reference \cite{R}
suggests an alternative approach to obtaining these equations.)
\vskip .3 cm
\noindent
\section{A difficult problem}
\vskip .2 cm
\noindent
As an illustration, consider the following mind boggling 
problem \cite{W}: 
\begin{figure}[t]
\begin{center}
\resizebox{!}{8.1 truecm}{\includegraphics{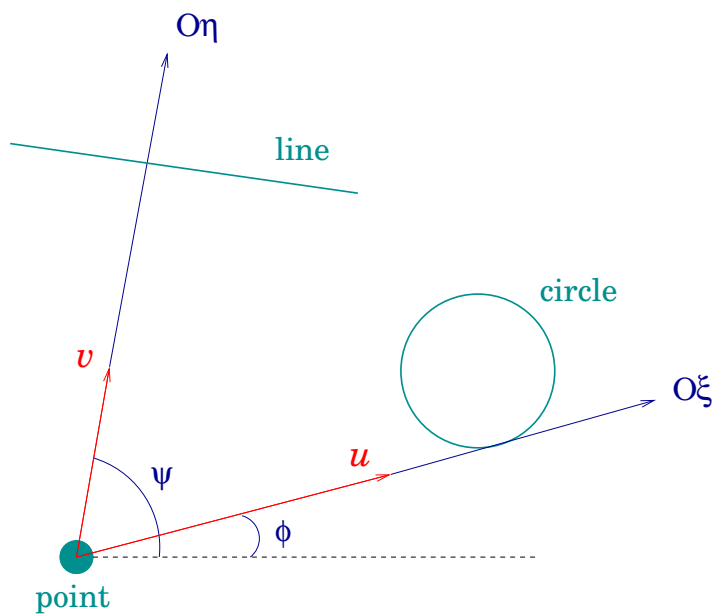}}
\vskip .2 cm
\parbox{13truecm}{\caption{\baselineskip=16 pt\small\label{oblique1}
An example of a difficult problem whose solution depends on the
kinematics of oblique axes.}
}
\end{center}
\end{figure}
\vskip .3 cm
\noindent
{\it A circle, a straight line, and a point lie in one
plane, and the position of the point is determined by the lengths $\tau$
of its tangent to the circle and $p$ of its perpendicular to the line.
Prove that, if the velocity of the point in made up of components 
$u$, $v$ in the directions of these lengths, and if their mutual
inclination is $\theta$, the component accelerations will be
\vskip .3 cm
\noindent
\[
\dot{u} - \frac{uv}{\tau}\cos{\theta},
\;\;\;\;\;\; \dot{v}+\frac{uv}{\tau}.
\]
\vskip .2 cm
\noindent
}
\begin{itemize}
\item[] {\bf Solution.} Take the axis $O\xi$ to be the tangent to the
circle, and $O\eta$ to be the axis perpendicular to the line. Set
$\theta=\psi-\phi$ and note that \underline{$\psi$ does not vary}
with time. It is then easy to check that equations (\ref{5}), with due
change in notation, reduce to the following set of equations
\[ 
a_{t}+a_{p}\cos\theta=\dot{u}+\dot{v}\cos\theta
\]
and
\be\label{90}
a_{p}+a_{t}\cos\theta=\dot{v}+\dot{u}\cos\theta+u\dot{\phi}\sin\theta .
\ee
\vskip .3 cm
\noindent
If we solve (\ref{90}) for $a_{t}$ and $a_{p}$ we obtain
\be \label{7}
a_{t}=\dot{u}-\frac{\cos\theta}{\sin\theta} u\dot{\phi} \;\;\;\;\;
\;\;\;\;
{\rm and} \;\;\;\;\;\;\;\;\; a_{p}=\dot{v}+\frac{u\dot{\phi}}{\sin\theta}.
\ee
\vskip .05 cm
\noindent
\begin{figure}[t]
\begin{center}
\resizebox{!}{7.6 truecm}{\includegraphics{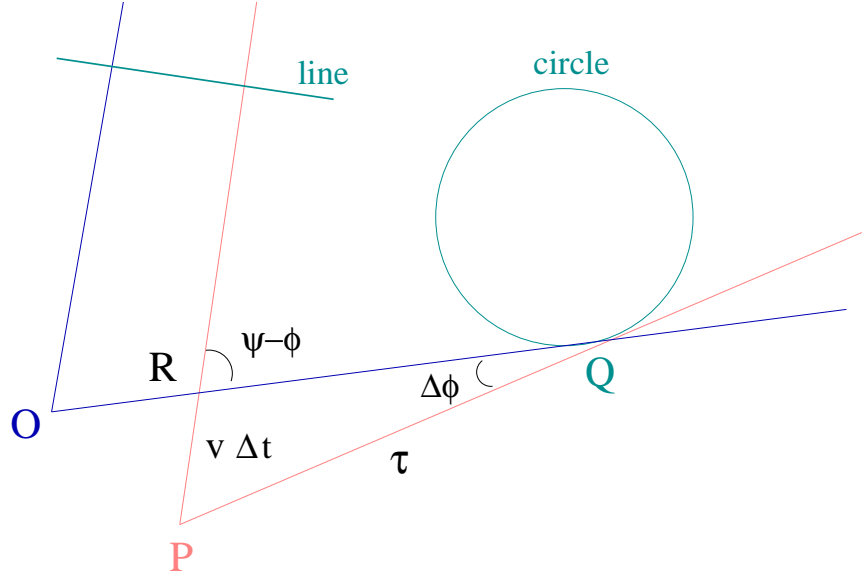}}
\vskip .2 cm
\parbox{13truecm}{\caption{\baselineskip=16 pt\small\label{oblique4}
The geometry of the illustrative problem. The particle moves from 
the point $O$ to the point $P$. Its position is determined by 
the tangent line to the circle and also by a perpendicular 
to a given line in the plane. The tangent lines at two different 
times meet at $Q$ and make an angle $\Delta \phi$.  }
}
\end{center}
\end{figure}
\noindent
To eliminate the variable $\dot{\phi}$ we need to consider  
the (messy) geometry of the problem. Let the two tangent lines to the
circle, drawn from the two positions of the particle at $O$ and $P$,
meet at a point $Q$, as shown in Fig. \ref{oblique4}. Note that the lines 
$OQ$ and $PQ$ form an angle $\Delta\phi$ with each other at $Q$.
Next, draw from the point $P$ a perpendicular to the given 
line. Let this perpendicular meet the line $OQ$ at a point $R$.
The perpendicular $PR$ makes an angle $\psi-\phi$ with $OQ$.
For small $\Delta t$, $P$ is near $O$, and 
we have, approximately, $PQ=\tau$ and $PR=v\Delta t$. The law of sines,
applied to the triangle $PQR$, gives
\be\label{6}
\frac{\Delta \phi}{v\Delta t} = \frac{\sin\theta}{\tau}
\;\;\;\;\;\;\;\;\; {\rm or} \;\;\;\;\;\;\;\;\; 
\dot{\phi}=\frac{v \sin\theta}{\tau}.
\ee
Substituting $\dot{\phi}$ given above in (\ref{7}), we obtain the desired
result.  
\end{itemize}
\subsection{Further examples}
\vskip .2 cm
\noindent
The equations for velocity and acceleration components on oblique 
axes can be used to provide solution to problems of motion in a
central force field when these problems are phrased as follows.
\vskip .2 cm
\noindent
\begin{itemize}
\item[{\bf 1.}] {\it A particle $P$ moves in a plane in such a way that
its velocity has two constant components $u$ and $v$, with $u$ parallel to
a fixed direction in the plane while $v$ is normal to a straight line from
the particle to a fixed point $O$ in the plane. Show that the acceleration
of the particle is directed along the line $OP$. (In fact, the particle
moves in a ellipse of eccentricity $u/v$, having $O$ as a focus.)}
\vskip .2 cm
\noindent
\item[{\bf 2.}] {\it A boat crosses a river with velocity of constant
magnitude $u$ always aimed toward a point $S$ on the opposite shore
directly across its starting position. The rivers also runs with uniform
velocity $u$. Compare this problem with the preceding one. (How far
downstream from $S$ does the boat reach the opposite shore?)
} 
\end{itemize}
\vskip .3 cm
\noindent
{\bf Acknowlgement.} {I was supported by Fapesp under grant  01/08485-6.}

\end{document}